\documentclass[aps,prb,superscriptaddress,preprint,showpacs]{revtex4}  
\usepackage{graphicx}
\usepackage{amsfonts}              
\usepackage{amssymb,amsmath}
\usepackage{bm}
\usepackage{comment}
\begin{document}
\title{Unified Description of the Intrinsic Spin-Hall Effects}
 \author{T. Fujita}       
   \affiliation{Information Storage Materials Laboratory, Electrical and Computer Engineering Department, National University of Singapore, 4 Engineering Drive 3, Singapore 117576}
    \affiliation{Data Storage Institute, A*STAR (Agency for Science, Technology and Research)
DSI Building, 5 Engineering Drive 1, Singapore 117608}
    \author{M. B. A. Jalil}
    \affiliation{Information Storage Materials Laboratory, Electrical and Computer Engineering Department, National University of Singapore, 4 Engineering Drive 3, Singapore 117576}
    \author{S. G. Tan}
    \affiliation{Data Storage Institute, A*STAR (Agency for Science, Technology and Research)
DSI Building, 5 Engineering Drive 1, Singapore 117608}
    \date{\today}    

\begin{abstract}
The intrinsic spin-Hall effects (SHE) in $p$-doped semiconductors [S.\ Murakami \emph{et al.}, Science {\bf 301}, 1348 (2003)] and two-dimensional electron gases with Rashba spin-orbit coupling [J.\ Sinova \emph{et al.}, Phys.\ Rev.\ Lett.\ {\bf 92}, 126603 (2004)] have been the subject of many theoretical studies, but their driving mechanisms have yet to be described in a unified manner. The former effect arises from the adiabatic topological curvature of momentum space, from which holes acquire a spin-dependent anomalous velocity. The SHE in the Rashba system, on the other hand, results from the momentum-dependent spin dynamics in the presence of an external electric field. The two effects clearly appear to originate from distinct mechanisms. Our motivation for this article is to address this apparent disparity and, in particular, to seek a unifying description of the effects. In this endeavor, we consider the explicit time-dependence of SHE systems starting with a general spin-orbit model. We find that by performing a gauge transformation of the general model with respect to time, a well-defined gauge field appears in time space which has the physical significance of an effective magnetic field. This magnetic field is shown to precisely account for the SHE in the Rashba system in the adiabatic limit. Remarkably, by carefully analyzing the equation of motion of the general model, this field component is also found to be the origin of the anomalous velocity due to the momentum space curvature. Our study therefore unifies the two seemingly disparate intrinsic SHEs under a common adiabatic framework.
\end{abstract}
\pacs{03.65.Vf, 03.65.-w,73.63.-b}
\maketitle
\section{Introduction}
The spin-Hall effects (SHE) are a family of phenomena in condensed matter systems in which an applied longitudinal electric field gives rise to a transverse spin current. The spin current arises from the transverse separation of spin species in the system, and the physics driving the separation mechanism can be rather distinct across different systems. The earliest prediction of transverse spin separation was made by D'yakonov and Perel' in the 1970's,\cite{dyakanov-perel} who studied the spin-dependent scattering mechanisms of carriers with localized impurities. The class of SHEs which occur as a result of the spin-orbit interaction of carriers with impurities are referred to in the literature as \emph{extrinsic}. Conversely, there are \emph{intrinsic} forms of the SHE, which have become an active field of research in more recent years,\cite{dai,kontani,shen,tan_prb} following two seminal papers: the first by Murakami \emph{et al.}\ which predicts a SHE of holes in $p$-doped semiconductors described in the Luttinger model,\cite{murakami} and the other by Sinova \emph{et al.}\ in two-dimensional electron gases (2DEG) formed in semiconductor heterostructures with Rashba spin-orbit coupling.\cite{sinova} Both effects are finite in the absence of disorder, and are characterized by a pure transverse spin current generated from the spin-orbit coupling (SOC) in the band structure of the system. Previous studies\cite{inoue,mishchenko} have shown that the SHE described in Ref.\ [\onlinecite{sinova}] for infinite Rashba systems vanishes when one includes vertex corrections to model the effects of impurity scattering. However, the effect may still be manifested in finite-sized systems\cite{mishchenko,xing,wang,adagideli} such as in mesoscopically confined 2DEGs [\onlinecite{xing}] (e.g.\ in quasi one-dimensional quantum wires), or in the presence of magnetic impurities.\cite{inoue2} In this work, we will analyze the SHE in Rashba systems without considering vertex corrections (i.e.\ in line with the original treatment in Ref.\  [\onlinecite{sinova}]). Instead, we will provide a phenomenological explanation of the vanishing SHE based on our own analysis later in the paper. In constrast, the SHE in the Luttinger hole system is robust to vertex corrections.\cite{murakami3}
\\
It is intriguing that the two SHEs in Refs.\ [\onlinecite{murakami}] and [\onlinecite{sinova}], although both intrinsic in nature, appear to originate from distinct mechanisms. The former effect in Ref.\ [\onlinecite{murakami}] is an adiabatic effect described via a
gauge potential which arises from the relaxation of hole spins to an
effective magnetic field in momentum $(\vec{k})$ space. The topological Berry curvature\cite{berry} of the gauge potential has the physical significance of a magnetic field in momentum space, and affects the trajectory of carriers in much the same way as a classical magnetic field does in real space. Here, the resulting magnetic Lorentz force in momentum space manifests itself as an additional (anomalous) velocity in real space. The semiclassical equations of motion of carriers in the presence of the Berry curvature have been derived previously,\cite{sundaram-niu} and will be revisited for the Luttinger system in  this article. It is found that the real space trajectory of holes along the transverse direction is spin-dependent, thus resulting in a finite SHE. The SHE in Rashba systems,\cite{sinova} on the other hand, was derived originally from a semiclassical analysis of electron spin dynamics in a Rashba 2DEG system, with no apparent relation to the $\vec{k}$-space topology. In the analysis,\cite{sinova} electrons were found to gain a momentum-dependent out-of-plane spin polarization in the presence of Rashba SOC and an external electric field, leading to a transverse separation of spins. It is often stated in the literature that the effect arises from the $\vec{k}$-anisotropic precession of spins.\cite{inoue,chen,engel,murakami3} As part of the motivation for this paper, we will clarify the mechanism and show that the effect is in fact an adiabatic effect in which spins become aligned to momentum-dependent effective magnetic fields.

From a heuristic viewpoint, the physical mechanism of the two SHEs [\onlinecite{murakami,sinova}] are clearly distinct: in the former, carriers acquire a spin-dependent anomalous velocity, whilst in the latter they acquire a momentum-dependent spin-polarization. Our motivation for this paper is two-fold. Firstly, we ask whether the SHE in Rashba systems can also be formulated within an adiabatic framework, and, secondly, whether the physical mechanisms of the two SHEs can be unified. In order to describe the Rashba SHE under an adiabatic formulation, it is instructive to make note of several points: (1) the adiabatic Berry curvature of momentum space of the Rashba system vanishes except as a $\delta$-function singularity at $\vec{k}=\vec{0}$.\cite{chang} Therefore, the spin-dependent anomalous velocity in the Rashba system vanishes for electrons with $\vec{k}\neq\vec{0}$ and thus does not contribute any transverse spin current. In contrast, the spin-Hall current in the Luttinger system results from the Dirac monopole curvature $(\sim \vec{k}/|\vec{k}|^3)$ of momentum space, (2) in the Rashba SHE, spins become tilted out-of-plane which appears contradictory to the adiabatic regime whereby they are assumed to follow perfectly the in-plane Rashba field, and (3) although the Berry curvature in the Rashba system exists only at a singular point in $\vec{k}$-space, the resulting Berry phase is finite, and previous studies have shown that the spin-Hall conductivity in the Rashba system is related to the Berry phase through the Kubo formula.\cite{shen,chen}

In this article, we find that the above remarks (1)-(3) can be consolidated into a consistent adiabatic theory that emerges from a consideration of the explicit time-dependence of SHE systems. In particular, a gauge field $\mathcal{A}_0(t)$ naturally appears in time space upon applying a unitary transformation to the system, which has the physical significance of a magnetic field in the transformed system. This magnetic field couples to the electron spin, and is shown to precisely account for the SHE in the Rashba system in the adiabatic limit; here, this limit amounts to the spins following the direction of the \emph{sum} of the Rashba field and the new effective field. Furthermore, the Berry phase can be equivalently expressed in terms of the adiabatic components of the gauge field $\mathcal{A}_0(t)$.
Thus, a gauge
field description can be attributed to both intrinsic SHEs,
although their respective gauge fields are defined in different
spaces (momentum and time).

Having identified that both intrinsic SHEs arise from gauge fields in the adiabatic limit, we finally embark on the problem of unifying the physical origin of the two effects. In the presence of an external electric field,
the momentum and time spaces become coupled through the usual drift
equation of charged carriers. Remarkably, by analyzing carefully the equations of motion of a general SOC
model, it is found that the anomalous velocity due to the Berry
curvature in momentum space is in fact a direct result of the effective magnetic
field component arising from $\mathcal{A}_0(t)$. In this sense,
the common origin of the two seemingly disparate SHEs is clarified.
\section{Theory}
\subsection{Carrier dynamics in the presence of Berry's curvature in momentum space in spin-orbit coupling systems}
\label{lutt.sec}
\subsubsection{Holes in the Luttinger system}
Let us briefly review the mechanism for the SHE of holes in $p$-doped semiconductors reported in Ref.\ [\onlinecite{murakami}]. The effective Luttinger Hamiltonian for holes in the valence band of conventional semiconductors is given by\cite{luttinger}
\begin{equation}
\hat{\mathcal{H}}_{\text{Lutt.}}=\frac{\hat{k}^2}{2}\left(\gamma_1 + \frac{5}{2}\gamma\right)-\gamma (\hat{k}\cdot\vec{S})^2+V(\hat{r}),
\label{lutt.eq}
\end{equation}
where $\gamma_1$, $\gamma$ are valence-band parameters defining the effective hole masses, $\hat{k}$ is the momentum operator, $\vec{S}$ is the vector of spin-$3/2$ matrices, and $V=V(\hat{r})$ is the potential energy (in our notation, a hat $(\hat o)$ signifies an
operator while an over-arrow $(\vec v)$ signifies a vector). The holes described by \eqref{lutt.eq} have a well-defined chirality,  $\hat{\lambda}=\hbar^{-1}\hat{k}\cdot\vec{S}/|\vec{k}|$. Because of the chirality-squared term in the Hamiltonian, states with opposite signs of  chirality ($\lambda=\pm 1/2$ and $\lambda=\pm 3/2$) are degenerate (they correspond to the light-hole and heavy-hole bands, respectively). In the presence of an external electric field $\vec{E}$, the potential energy term is $V(\hat{r})=e\vec{E}\cdot\hat{r}$, where $-e$ is the electron charge. Parameterizing the momentum vector $\vec{k}=|\vec{k}|(\sin\theta\cos\phi,\sin\theta\sin\phi,\cos\theta)$, we proceed to define a $4\times 4$ unitary matrix $U(\vec{k})$,
\begin{equation}
U(\vec{k}) = \exp{\left(i\theta(\vec{k})S_y\right)}\cdot\exp{\left(i\phi(\vec{k})S_z\right)},
\label{umat.eq}
\end{equation}
which aligns the reference spin axis to be along the direction of $\vec{k}$, i.e.\ it satisfies the relation $U(\vec{k})\left(\vec{k}\cdot\vec{S}\right)U^\dagger (\vec{k}) = |\vec{k}| S_z$. The effective, diagonalized Hamiltonian $\hat{\mathcal{H}}'_{\text{Lutt.}}=U(\vec{k})\hat{\mathcal{H}}_{\text{Lutt.}}U^\dagger(\vec{k})$ then reads
\begin{equation}
\hat{\mathcal{H}}'_{\text{Lutt.}}=\frac{\hat{k}^2}{2}\left(\gamma_1 + \frac{5}{2}\gamma\right)-\gamma |\hat{k}|^2 S_z^2+U(\vec{k})V(\hat{r})U^\dagger(\vec{k}).
\label{luttdiag.eq}
\end{equation}
In the last term of \eqref{luttdiag.eq}, the position operator $\hat{r}=i\partial_k$ acts as a partial derivative in momentum space, and we obtain from the $\vec{k}$-dependence of $U$:
\begin{equation}
U(\vec{k}) \left(e\vec{E}\cdot i\partial_k\right) U^\dagger (\vec{k}) =e\vec{E}\cdot\left(\hat{r}+iU (\vec{k}) \partial_k U^\dagger (\vec{k})\right).
\label{covariant.eq}
\end{equation}
Thus, under the local transformation, the position operator transforms into covariant form, $\hat{r}\rightarrow \hat{R} = \hat{r}-\mathcal{A}(\vec{k})$, where $\mathcal{A}(\vec{k})\equiv -i U(\vec{k}) \partial_k U^\dagger(\vec{k})$ is a gauge field in reciprocal space. 
Thus far, the transformation $\hat{\mathcal{H}}_{\text{Lutt.}}\rightarrow\hat{\mathcal{H}}'_{\text{Lutt.}}$ is exact. Being a pure gauge field, $\mathcal{A}(\vec{k})$ induced by the transformation has no associated curvature. Assuming adiabatic transport, in which we neglect mixing between the light-hole and heavy-hole bands, and applying an Abelian approximation within each hole band, we are left with only the diagonal gauge field components of the respective $2\times 2$ hole band subspaces. Explicitly, the Abelian gauge fields are given by
\begin{equation}
\mathcal{A}^{\text{ad.}}(\vec{k})=-\lambda\cos\theta\nabla_\mathbf{k} \phi,
\end{equation}
where the superscript \emph{ad.}\ denotes adiabatic transport, and $\lambda$ is the hole chirality. The corresponding gauge invariant quantity (which is thus related to a real physical effect) is the curvature tensor $\Omega(\vec{k})$, defined by
\begin{equation}
\Omega_k(\vec{k}) = \partial_{k_i} \mathcal{A}^{\text{ad.}}_{k_j}(\vec{k})-\partial_{k_j} \mathcal{A}^{\text{ad.}}_{k_i}(\vec{k}).
\label{curvature.eq}
\end{equation}
The curvature $\Omega(\vec{k})$ above is frequently called the Berry curvature in momentum space. In the present case, a simple calculation reveals that the Berry curvature is
\begin{equation}
\Omega(\vec{k})= \lambda\frac{\vec{k}}{|\vec{k}|^3},
\label{luttcurv.eq}
\end{equation}
i.e.\ it is a Dirac monopole with strength $eg=\lambda$. It turns out that the $\vec{k}$-space curvature \eqref{luttcurv.eq} has important implications on carrier dynamics. In particular, $\Omega(\vec{k})$ can be regarded as a magnetic field in $\vec{k}$-space, which gives rise to a $\vec{k}$-space Lorentz-type force. The modified semiclassical equations of motion for carriers in the presence of a non-trivial curvature in $\vec{k}$-space have been derived elsewhere to be:\cite{sundaram-niu}
\begin{eqnarray}
\hbar \dot{\vec{k}} &=& -e\vec{E},\\
\dot{\vec{r}} &=& \frac{1}{\hbar}\nabla_\mathbf{k} \epsilon-\dot{\vec{k}}\times\Omega(\vec{k}), \label{eom.eq}
\end{eqnarray}
where the over-dot signifies time differentiation, and $\epsilon$ is the energy eigenvalue of the system. The final term in \eqref{eom.eq} is the Lorentz-type force in $\vec{k}$-space, and is equivalent to an additional velocity of electrons corresponding to the so-called anomalous Karplus-Luttinger term.\cite{karp-lutt} Substituting the expression for the curvature \eqref{luttcurv.eq} into the equation of motion, the anomalous velocity component is given by 
\begin{equation}
v^{\text{anom.}}=- \lambda \dot{\vec{k}}\times \frac{\vec{k}}{|\vec{k}|^3},
\label{luttanom.eq}
\end{equation}
which is perpendicular to both the applied electric field $\vec{E}$ and $\vec{k}$. Since the chirality of the holes has sign $\lambda>0$ $(<0)$ for hole spins (anti-)parallel to the electron momentum, the anomalous velocity is also perpendicular to the spin $\vec{S}$, and points along opposite directions depending on the sign of the chirality. This transverse separation of the spins gives rise to the SHE of holes in the Luttinger system.
\subsubsection{Conduction electrons in the Rashba system}
We now analyze the Berry curvature in momentum space for the case of the linear Rashba SOC,\cite{rashba1,rashba2} which is present in two-dimensional electron gases formed in semiconductor heterostructures. We begin with the generalized spin-orbit Hamiltonian,
\begin{equation}
\hat{\mathcal{H}}=\frac{\hat{p}^{\text{ }2}}{2m}-\gamma\vec{\sigma}\cdot\vec{B}(\hat{k})+V(\hat{r}),
\label{sochamil.eq}
\end{equation}
where $m$ is the effective electron mass, $\gamma$ is the SOC strength, $\vec{\sigma}=\{{\sigma}_i\}$ is the vector of Pauli spin matrices, $\vec{B}(\hat{k})$ is a momentum-dependent effective magnetic field, and $V(\hat{r})=e\vec{E}\cdot\hat{r}$ in the
presence of an external electric field $\vec{E}$. The above Hamiltonian captures the physics of many other types of SOC, including the linear and cubic Dresselhaus\cite{dressel} and strain-induced\cite{malshukov,jiang} SOC systems. The Luttinger Hamiltonian \eqref{lutt.eq} can also be transformed to be of this general form when re-cast in terms of the SO(5) Clifford algebra as was done in Ref.\ [\onlinecite{murakami2}], although this representation is in a 5-dimensional space rather than the usual spin-$\frac{1}{2}$ space. The single particle eigenstates of the Hamiltonian are of the form $|\psi_\pm\rangle = \exp{\left( i\vec{k}\cdot\vec{r} \right)}\chi_\pm(\vec{k})$, i.e.\ a product of the spatial plane wave
state and the spinor part which encodes the electron spin state. For any $\vec{k}$, the spin degeneracy is lifted between the two eigenstates $|\psi_\pm\rangle$, which have corresponding spin-orbit energy eigenvalues of $\epsilon_\pm = \pm \gamma |\vec{B}(\vec{k})|$. 
Let us rotate the reference spin axis such that
it points along the direction of the spin-orbit field
$\vec{B}(\vec{k})$, i.e. we diagonalize the Hamiltonian with respect
to $\vec{B}(\vec{k})$.  By parameterizing the spin-orbit field in terms of spherical angles, $\vec{B}=|\vec{B}|(\sin\theta\cos\phi,\sin\theta\sin\phi,\cos\theta)\equiv|\vec{B}|\vec{n}$, where $\theta$ and $\phi$ are explicit functions of $\vec{k}$, the diagonalization may be achieved through the SU(2) rotation matrix $U=U(\vec{k})$ given by Eq.\ \eqref{umat.eq} but with the replacements $S_y\rightarrow \sigma_y /2$, $S_z \rightarrow \sigma_z/2$. However, the choice of $U$ for the diagonalization is not unique: for convenience we shall adopt another rotation matrix given by\cite{tatara}
\begin{equation}
U(\vec{k})=\vec{m}(\vec{k})\cdot\vec{\sigma},
\label{Umatrix.eq}
\end{equation}
where $\vec{m}=\left(\sin{\frac{\theta}{2}}\cos{\phi},\sin{\frac{\theta}{2}}\sin{\phi},\cos{\frac{\theta}{2}}\right)$.
The effective, diagonalized Hamiltonian is given by
\begin{equation}
\hat{\mathcal{H}}'= U \hat{\mathcal{H}} U^\dagger=\frac{\hat{p}^{\text{ }2}}{2m}-\gamma{\sigma_z}|\vec{B}(\hat{k})|+UV(\hat{r})U^{\dagger}.
\label{diaglhs.eq}
\end{equation}
The $\sigma_z$ Pauli matrix in the diagonalized spin-orbit term represents the two spin states either parallel (the ground state) or anti-parallel (the $\epsilon_+$ state) to the spin-orbit field $\vec{B}(\vec{k})$. In the last term of \eqref{diaglhs.eq}, the position operator $\hat{r}=i\partial_k$ is transformed into the covariant form of Eq.\ \eqref{covariant.eq}, i.e.\ $\hat{r}\rightarrow \hat{R} = \hat{r}-\mathcal{A}(\vec{k})$, where $\mathcal{A}(\vec{k})$ is a SU(2) gauge field in reciprocal space. From Eqs.\ \eqref{covariant.eq} and \eqref{Umatrix.eq}, the gauge
field components can be represented in terms of the $\vec{m}$-vector
and the Pauli spin matrices, i.e.
\begin{equation}
\mathcal{A}_{k_i}(\vec{k}) = \left(\vec{m}\times\partial_{k_i}\vec{m}\right)\cdot\vec{\sigma}=\vec{A}_{k_i}\cdot\vec{\sigma},
\label{ak.eq}
\end{equation}
and in terms of the $\vec{n}$-vector by replacing $\vec{A}_{k_i}$ in the above equation with
\begin{equation}
\vec{A}_{k_i} = \frac{1}{2}\left(\vec{n}\times\partial_{k_i}\vec{n}\right)+(\vec{A}_{k_i}\cdot\vec{n})\vec{n},
\label{akn.eq}
\end{equation}
where $i=x,y,z$ are real space coordinates.
Up to this point, the transformation of the Hamiltonian is general.
We now impose the adiabatic approximation, in which mixing between
the two eigenstates of the diagonalized Hamiltonian is neglected.
Mathematically, this corresponds to retaining only the diagonal terms of $\mathcal{A}(\vec{k})$, i.e.\ the $\sigma_z$ coefficients in Eq.\ \eqref{ak.eq}, from which we obtain an Abelian gauge field known as the Berry connection, $\mathcal{A}^{\text{ad.}}(\vec{k},s)$. $\mathcal{A}^{\text{ad.}}(\vec{k},s)$ has two values, representing
the two spin states, $s=\pm 1$, of the diagonalized Hamiltonian (we
denote the ground state as $s=+1$), and which correspond to the
diagonal terms of $\mathcal{A}_{k_i}(\vec{k})$. Explicitly, the
Abelian gauge field is given by
\begin{equation}
\mathcal{A}^{\text{ad.}}(\vec{k},s)=-\frac{s}{2}\left(1-\cos\theta\right)\nabla_\mathbf{k} \phi.
\label{adiab.eq}
\end{equation}
The curvature tensor $\Omega(\vec{k})$ of this connection, defined by Eq.\ \eqref{curvature.eq}, is invariant with respect to the gauge transformation $U(\vec{k})$. From the definition, it is clear that $\Omega(\vec{k})$ respects the same symmetry in $s$ as the connection, i.e.\ $\Omega(\vec{k},s)=-\Omega(\vec{k},-s)$. 

In principle, one can define the curvature $\Omega$ in any arbitrary
space. For example, in the special case of the magnetic field space $\vec{B}$, the Berry curvature has the classic form of Dirac's monopole,\cite{berry}
\begin{equation}
\Omega(\vec{B})= s\frac{\vec{B}}{2|\vec{B}|^3}.
\end{equation}
The above relation is general and applies to any SOC system. One can
transform the curvature from $\vec{B}$-space to any other space (e.g.\
$\vec{k}$-space) by using the relation:\cite{bliokh,tan-upub}
\begin{equation}
\Omega_{k}(\vec{k})=\epsilon_{ijk}\Omega(\vec{B})\cdot\left( \frac{\partial \vec{B}}{\partial k_i}\times\frac{\partial \vec{B}}{\partial k_j}\right),
\label{B2k-curv.eq}
\end{equation}
where $\epsilon_{ijk}$ is the Levi-Civita symbol. Generally, the curvature in momentum space $\Omega(\vec{k})$ is not
the Dirac monopole field (although it still is for the case
of the Luttinger Hamiltonian). The actual form of $\Omega(\vec{k})$
depends on the $\vec{k}$-dependence of the effective magnetic
field. We saw in Eq.\ \eqref{eom.eq} how this curvature $\Omega(\vec{k})$ gave rise to an anomalous velocity which resulted in the SHE in $p$-doped semiconductors. However, the same reasoning cannot be applied to the SHE in the Rashba system, as $\Omega(\vec{k})$ in this system is vanishing (for $\vec{k}\neq \vec{0}$) as we outline below.\\
The Hamiltonian in the presence of the Rashba SOC is given by\cite{rashba1,rashba2}
\begin{equation}
\hat{\mathcal{H}}_R=\frac{\hat{p}^{\text{ }2}}{2m}+\alpha\left(\hat{k}_x {\sigma}_y - \hat{k}_y {\sigma}_x\right),
\label{rashbahamil.eq}
\end{equation}
where $\alpha$ is the Rashba spin-orbit coupling parameter expressed in units of eVm. The effective magnetic field is given by $\vec{B}_R(\hat{k})=(\hat{k}_y,-\hat{k}_x)$, and the eigenvectors are $|\vec{k},\pm\rangle = 1/\sqrt{2} \exp\left(i\vec{k}\cdot\vec{r}\right)\left(\mp i k^{-1}(k_x-ik_y),1\right)^\text{T}$, with the corresponding energy eigenvalues of $\epsilon_\pm = \pm \alpha k$ where $k = |\vec{k}|=\sqrt{k_x^2+k_y^2}$ is the in-plane wave-vector magnitude. In momentum space, the effective magnetic field $\vec{B}_R$ is directed along $\theta=\pi/2$
and $\phi=\tan^{-1} (-k_x/k_y)$, and from Eq.\ \eqref{adiab.eq} the
Berry connection is given by $\mathcal{A}^{\text{ad.}}(\vec{k},\pm)= \pm \frac{1}{2k ^{2}}(-k_y,k_x,0)$. Evidently, the curvature \eqref{curvature.eq} of this connection  is trivial, i.e.\ $\Omega(\vec{k})=\vec{0}$ over the entire $\vec{k}$-space, except at the singularity point at $\vec{k}=\vec{0}$ where the $k_z$-component of the curvature has non-vanishing value $\pm \pi$, i.e.\ the curvature is of the form $\Omega(\vec{k},\pm)=(0,0,\pm\pi
\delta(\vec{k}))$.\cite{chang} Thus, conduction electrons having a finite momentum in the 2DEG plane do not experience any Lorentz-type force in $\vec{k}$-space, as is the case for holes in the Luttinger system. Furthermore, even if this force existed, it would only separate the Rashba SOC eigenstates (whose spins lie entirely in-plane) in the transverse direction, and thus cannot explain the out-of-plane spin polarization acquired by the electrons in the SHE. Now the question arises as to whether the SHE in
Rashba systems can be described within a gauge field framework. In
particular, the out-of-plane spin polarization seems to suggest the
presence of an additional magnetic field in the system. It turns out
that such a gauge formulation does exist, but one must turn to
another parameter space, namely the time space.
\subsection{Time component of the gauge field in spin-orbit coupling systems with an electric field}
When considering the temporal evolution of a quantum system, the unitary transformation is explicitly time-dependent, i.e.\ $U=U(t)$. In SHE systems, the $t$-dependence of the unitary transformations naturally arises due to the acceleration of carriers in the presence of an electric field: the electron wave-vector $\langle\hat{k}\rangle$ changes linearly in $t$, and consequently $\langle\vec{B}(\hat{k})\rangle$ acquires a time-dependence. To incorporate the explicit time-dependence of the system quantum mechanically, we switch to the interaction picture.\cite{townsend} In this picture, the original Hamiltonian \eqref{sochamil.eq} is split into two parts, $\hat{\mathcal{H}} = \hat{\mathcal{H}}_0 + \hat{\mathcal{H}}_1$, where 
\begin{equation}
\hat{\mathcal{H}}_0 = e \vec{E}\cdot\hat{r}
\end{equation}
governs the time evolution of the operators, and 
\begin{equation}
\hat{\mathcal{H}}_1 = \frac{\hat{p}^{\text{ }2}}{2m} -\gamma\vec{\sigma}\cdot\vec{B}(\hat{k})
\end{equation}
governs the time evolution of the states. In the usual sense, an operator $\hat{A}$ in the Schr\"{o}dinger picture is transformed to the interaction picture (subscript $I$) as $\hat{A}_I(t) = e^{i \hat{\mathcal{H}}_0 t/\hbar} \hat{A} e^{-i \hat{\mathcal{H}}_0 t/\hbar}$ and carries an explicit time-dependence by satisfying the Heisenberg relation, $\dot{\hat{A}}_I=(i\hbar)^{-1} [\hat{A}_I,\hat{\mathcal{H}}_0]$. In particular, the momentum operator in the new picture is found to be $\hat{p}_I(t) = \hat{p}-e\vec{E}t$, i.e.\ with the expected linear time-dependence due to the electric field. The state vectors $|\psi(t)\rangle$ in the Schr\"{o}dinger picture correspondingly transform as $|\psi_I(t)\rangle = e^{i \hat{\mathcal{H}}_0 t/\hbar} |\psi(t)\rangle$, and evolve according to the new ``Schr\"{o}dinger equation'',
\begin{equation}
\hat{\mathcal{H}}_I(t) |\psi_I(t)\rangle = i\hbar \partial_t |\psi_I(t)\rangle,
\label{schro.eq}
\end{equation}
where $\hat{\mathcal{H}}_I(t) = e^{i \hat{\mathcal{H}}_0 t/\hbar} \hat{\mathcal{H}}_1 e^{-i \hat{\mathcal{H}}_0 t/\hbar}$. For the
case of linear (e.g.\ Rashba) spin-orbit coupling, $\hat{\mathcal H}_I (t)$ is evaluated to be
\begin{equation}
\hat{\mathcal{H}}_I(t) =\frac{\hat{p}_I^{\text{ }2}}{2m} -\gamma \vec{\sigma}\cdot\left( \vec{B}(\hat{k}) - \frac{e E_i t}{\hbar} \frac{\partial \vec{B}(\hat{k})}{\partial k_i} \right)\equiv\frac{\hat{p}_I^{\text{ }2}}{2m}-\gamma\vec{\sigma}\cdot \vec{B}(t).
\label{interham.eq}
\end{equation}
Higher order spin-orbit terms ($\sim k^n,n \geq 2$) generally lead to correspondingly higher order partial derivatives of the spin-orbit field in $\hat{\mathcal{H}}_I(t)$. The Hamiltonian \eqref{interham.eq} governing the state vector evolution in the interaction picture is that of an electron subject to an explicitly time-dependent magnetic field, which we denote as $\vec{B}(t)$. Analogous to our previous treatment, we proceed to diagonalize the Schr\"{o}dinger equation \eqref{schro.eq} at time $t$, by applying a unitary rotation $U(t)$ (defined as in Eq.\ \eqref{Umatrix.eq} but with $\theta$ and $\phi$ in $\vec{m}$ carrying an explicit time-dependence). This transformation aligns the $\tilde{z}$-axis to be parallel to the instantaneous magnetic field $\vec{B}(t)$, i.e.
\begin{eqnarray}
U(t)\hat{\mathcal{H}}_I(t) U^\dagger(t)&=& U(t) \left(i\hbar \partial_t \right) U^\dagger(t),\nonumber\\
\frac{\hat{p}_I^{\text{ }2}}{2m}-\gamma \sigma_z |\vec{B}(t)| &=& i\hbar\partial_t+i\hbar U(t) \partial_t U^\dagger(t),\nonumber\\
 &\equiv& \hat{\epsilon}-\hbar\mathcal{A}_0(t).
 \label{hamil_t_eff.eq}
\end{eqnarray}
where $\hat{\epsilon}=i\hbar\partial_t$ is the energy operator. On the left-hand-side, the local transformation diagonalizes the time-dependent Zeeman term as required. On the right-hand-side, we obtain from the time-dependence of $U$ a gauge field $\mathcal{A}_0(t)\equiv -i U(t) \partial_t U^\dagger (t)$ related to the temporal evolution of the system. From the relations in Eqs.\ \eqref{ak.eq} and \eqref{akn.eq}, we can express the gauge field as $\mathcal{A}_0(t)=\vec{A}_t\cdot\vec{\sigma}$, where $\vec{A}_t=\vec{m}\times\dot{\vec{m}}=\frac{1}{2}\vec{n}\times\dot{\vec{n}}+(\vec{A}_t\cdot\vec{n})\vec{n}$. Thus, the term $\hbar\mathcal{A}_0(t)$ represents an additional Zeeman-like term, indicating the presence of an effective magnetic field in the rotating frame. We elucidate the origin of this field  in more detail below. 
%
%
The unitary transformation $U(t)$ we invoked defines an instantaneous angular velocity $\vec{\omega}^l=\vec{\omega}^l(t)$ of the coordinates (in the laboratory frame, $l$) as it follows the time-dependent magnetic field. In the rotating frame $r$, this vector is given by $\vec{\omega}^r$, where $\vec{\sigma}\cdot\vec{\omega}^r = U \left( \vec{\sigma}\cdot\vec{\omega}^l \right) U^\dagger$. Since $\dot{U}=\frac{1}{2}i U \vec{\sigma}\cdot\vec{\omega}^l$ [\onlinecite{apoorva}], the Zeeman-like term $\hbar\mathcal{A}_0(t)$ in \eqref{hamil_t_eff.eq} thus yields $-\hbar/2 (\vec{\sigma}\cdot\vec{\omega}^r)$, which corresponds to an effective magnetic field $-\vec{\omega}^r$ (omitting a scaling factor) in the rotating frame. This translates into an effective magnetic field $\vec{B}_t=-\vec\omega^l$ in the laboratory frame. If we now denote by $\vec{n}=\vec{n}(t)$ the unit vector pointing along the direction of the magnetic field at time $t$, we have the equation of motion $\dot{\vec{n}} = \vec{\omega}^l\times\vec{n}$. Performing a post cross product on both sides by $\vec{n}$, one arrives at the expression for the angular velocity $\vec{\omega}^l = \vec{n}\times\dot{\vec{n}}+\left(\vec{\omega}^l\cdot\vec{n}\right)\vec{n}$, or, in terms of the effective magnetic field,
\begin{equation}
\vec{B}_t = \dot{\vec{n}}\times\vec{n}+\left(\vec{B}_t\cdot\vec{n}\right)\vec{n}.
\label{btt.eq}
\end{equation}
 Thus, the effective magnetic field arising from the gauge field $\mathcal{A}_0(t)$ of the unitary transformation has a component along $\dot{\vec{n}}\times\vec{n}$ and along $\vec{n}$. Note that it does not have any component along $\dot{\vec{n}}$. As we noted previously, the unitary rotation matrix used by us \eqref{Umatrix.eq} is not unique. Specifically, different rotation matrices $U_i$, each specifying distinct angular velocities $\vec{\omega}_i ^l$, can be used to align the reference $\tilde{z}$-axis with the instantaneous magnetic field $\vec{B}(t)$; the freedom of choice here lies in determining the trajectory of the remaining $\tilde{x},\tilde{y}$-axes. The second term on the right-hand-side of Eq.\ \eqref{btt.eq} reflects the particular choice of the gauge transformation $U_i$. It is not an invariant of the gauge transformation (its magnitude being dependent on the particular gauge choice), and does not represent a physical field. However, the first component $\dot{\vec{n}}\times\vec{n}$ of the effective magnetic field is invariant with respect to the gauge transformation, depending only on the time-dependence of the magnetic field $\vec{B}(t)$. This term can be understood to be a direct consequence of the time-dependent rotation of the axes.\cite{stern,xiao} The same expression can be derived classically by directly comparing the spin vector in adjacent time frames\cite{stern}---as a complement to the quantum derivation, the classical treatment is shown in detail in the Appendix. The $\dot{\vec{n}}\times\vec{n}$ component represents a physical magnetic field which couples to the electron spins,\cite{stern,xiao} and, as we show below, is precisely the component which leads to the SHE in Rashba 2DEG systems.
\section{Analysis}
\subsection{The intrinsic SHE due to Rashba SOC}
The Hamiltonian of conduction electrons in the Rashba system is given by Eq.\ \eqref{rashbahamil.eq}. Following our analysis above, the time-dependence of the effective Rashba field $\vec{B}_R$ due to the electrons' motion in momentum space necessarily gives rise to a secondary component $\vec{B}_\bot=\dot{\vec{n}}\times\vec{n}$, where $\vec{n}=p^{-1}(p_y,-p_x,0)$ is the unit vector in the direction of $\vec{B}_R$. We assume a longitudinally applied electric field along the $\tilde{x}$-direction $\vec{E}=E_x \tilde{x}$, so that $\dot{\vec{n}}=p^{-1}(0,eE_x,0)$. Because $\vec{B}_R$ is strictly in-plane (i.e.\ it lies in the $\tilde{x},\tilde{y}$-plane of the 2DEG), the  term  $\dot{\vec{n}}\times\vec{n}$ represents an out-of-plane magnetic field component  which is along the $\tilde{z}$-direction by convention. Next, we apply the adiabatic condition for the electron spins. In the ideal adiabatic limit, the magnetic field $|\vec{B}_R|$ is infinitely strong, so the spins always remain aligned to it as it varies with time. In reality, $|\vec{B}_R|$ is finite and there is a non-zero secondary component $\vec{B}_\bot$, and the relevant condition is $|\vec{B}_R| \gg |\vec{B}_\bot |$, i.e.\ the electron spin is primarily aligned to $\vec{B}_R$, but with a small deviation along $\vec{B}_\bot$. In terms of the parameters of the Rashba system, the adiabatic condition reads as
\begin{equation}
\frac{\alpha  k^2 |e|}{e} \gg E_x.
\end{equation}
Inserting typical values for the Rashba parameter $\alpha = 10^{-11}$ eVm and the Fermi wave-vector $k = 10^{8}$ m$^{-1}$, we arrive at the condition $E_x \ll \sim 10^{5}$ Vm$^{-1}$, which usually holds true in experiments. Assuming that the spin of electrons follow the direction of the net effective magnetic field, $\vec{B}_\Sigma$, which is the sum of the spin-orbit field $\vec{B}_R$ and the secondary component, the classical spin vector is given by
\begin{equation}
\vec{s}=\pm \frac{\hbar}{2} \frac{\vec{B}_\Sigma}{|\vec{B}_\Sigma|},
\end{equation}
 where $\pm$ represents spin aligned parallel $(+)$ or anti-parallel $(-)$ to the net field. To first order, the component of the spin along the $\tilde{z}$-direction is
\begin{equation}
s_z = \pm\frac{1}{|\vec{B}_\Sigma|}\frac{\hbar}{2}\left(\dot{\vec{n}}\times\vec{n}\right)\cdot\tilde{z},
\label{sz.eq}
\end{equation}
where, to be consistent in units, the magnetic field in the denominator is defined in terms of its equivalent angular velocity. Note that in the convention above the $+$ corresponds to the ground state $\epsilon_-$, whilst $-$ corresponds to the eigenstate $\epsilon_+$. In the adiabatic limit, the magnitude of $\vec{B}_\Sigma$ approaches that of $\vec{B}_R$, and applying this limit to Eq.\ \eqref{sz.eq}, we obtain for the out-of-plane spin polarization
\begin{eqnarray}
s_z & \approx & \pm\frac{1}{|\vec{B}_R|}\frac{\hbar}{2}\left(\dot{\vec{n}}\times\vec{n}\right)\cdot\tilde{z}, \nonumber \\ 
&=& \pm\frac{\hbar^2}{2\alpha p}\frac{\hbar}{2}\left( -\frac{1}{p^2} e E_x p_y\right), \nonumber \\
&=& \mp\frac{e \hbar^3 p_y E_x}{4 \alpha p^3}.
\label{szrashba.eq}
\end{eqnarray}
Eq.\ \eqref{szrashba.eq} above describes a transverse separation of spins in the Rashba system. Let us consider the case for the ground state. Since the spin $s_z \propto - p_y$, we find that electrons moving in the $+\tilde{y}$-direction are polarized out-of-plane along the $-\tilde{z}$ direction, whereas those moving in the $-\tilde{y}$-direction are polarized along $+\tilde{z}$. For the other eigenstate, the direction of the polarization is reversed, and hence there is a certain degree of canceling of the polarization if both eigenstates are present. However, at the Fermi level, there are more electrons in the ground state $\epsilon_-$, giving rise to a net transverse spin separation and hence the SHE described in Ref.\ \onlinecite{sinova} (see Eqs.\ (5)$-$(7) there). Summing over the Fermi surfaces of the two eigenstates yields an intrinsic spin-Hall (sH) conductivity of $\sigma_{\text{sH}}\equiv j_y^z / E_x =-e/8\pi$, where $j_y^z=\hbar/4 \{ s_z,v_y\}$ is the transverse spin-current. From our analysis above, we have clarified that the SHE in Rashba systems occurs as a result of an adiabatic process, in which electrons' spins become aligned to momentum-dependent magnetic fields that arise from the time-dependence of the system. The effect is therefore \emph{not} due to the precessional behavior of spins, as is often stated in the literature.
\subsubsection{Berry's phase}
We alluded earlier to previous work which related the intrinsic spin-Hall conductivity in Rashba systems to the $\vec{k}$-space Berry phase of electrons through the Kubo formula.\cite{shen} It was found there that $\sigma_{\text{sH}} = e\varphi_{\pm}/8\pi^2$, where $\varphi_{\pm}$ is the Berry phase of electrons,
\begin{equation}
\varphi_\pm = \oint \mathcal{A}_0^{\text{ad.}}(\vec{k})\cdot\vec{dk}=-\frac{s}{2}\oint(1-\cos\theta)\nabla_\mathbf{k} \phi \cdot\vec{dk}.
\end{equation}
The natural parameterization for the vector $\vec{k}$ is the time variable $t$, and rewriting the line integral above in terms of $t$ we obtain
\begin{eqnarray}
\varphi_\pm &=& -\frac{s}{2}\int(1-\cos\theta)\nabla_\mathbf{k} \phi\cdot\dot{\vec{k}} {dt}, \nonumber \\
&=&-\frac{s}{2}\int(1-\cos\theta)\dot{\phi} {dt}, \nonumber \\
&\equiv &\int \mathcal{A}_0^{\text{ad.}}(t) {dt}.
\end{eqnarray}
Thus the Berry phase and hence the intrinsic spin-Hall conductivity of the Rashba system can be written equivalently in terms of the time component of the adiabatic gauge field, $\mathcal{A}_0^{\text{ad.}}(t)$.
\subsubsection{Effects of disorder}
Previous studies have shown that the intrinsic SHE in infinite Rashba systems vanishes in the presence of disorder.\cite{inoue,mishchenko} Specifically, the vertex correction was shown to exactly cancel the intrinsic conductivity of $e/8\pi$ even in the weak scattering limit. We provide a heuristic argument based on our analysis for the vanishing SHE. In the presence of disorder, the scattering provides a braking effect which cancels the acceleration of carriers on average in the steady state.\cite{adagideli} This implies that in the steady state we have $\langle \dot{\vec{k}} \rangle = 0$, i.e.\ there is no net change in the momentum and thus the magnetic field component $\vec{B}_\bot=\dot{\vec{n}}\times\vec{n}$ averages out to zero. Note, however, that this picture is an oversimplification,\footnote{This argument was stated previously in the context of spin precession in [\onlinecite{murakami3}]: in the presence of impurities, the scattering scrambles the spin precession sufficiently such that no net SHE results.} and that the SHE in Rashba systems does not vanish in general. For example, the SHE persists in finite-sized systems\cite{mishchenko,xing,wang,adagideli} and in the presence of spin-dependent impurities.\cite{inoue2}
\\
\subsection{The intrinsic SHE due to linear Dresselhaus SOC}
The case for the linear Dresselhaus spin-orbit coupling is also easily verified by our analysis. The Dresselhaus spin-orbit Hamiltonian is given by
\begin{equation}
\mathcal{H}_D = \beta\left(\hat{k_y} \sigma_y-\hat{k_x} \sigma_x\right)\equiv -\beta\vec{\sigma}\cdot \vec{B}_D,
\end{equation}
where $\beta$ is the Dresselhaus SOC strength and $\vec{B}_D$ is the effective Dresselhaus SOC field. Here we have $\vec{n}=p^{-1}(p_x,-p_y,0)$, and we find that $(\dot{\vec{n}}\times\vec{n})_z=+e E_x p_y/p^2$. Consequently, the out-of-plane spin polarization $s_z$ has the same magnitude but opposite sign compared to the Rashba SOC case. This is in agreement with previous theoretical studies\cite{shen} which predicts the spin-Hall conductivity in this system to be $\sigma_{\text{sH}}=e/8\pi$.
\section{Discussions}
Having established the Rashba SHE as an adiabatic effect, we now identify two common traits of the two intrinsic SHEs; adiabaticity and time-dependence. For the Rashba system, we found that the time-dependent spin-orbit field $\vec{B}_R(t)=|\vec{B}_R(t)|\vec{n}$ is always accompanied by an additional effective magnetic field, $\vec{B}_\bot=\dot{\vec{n}}\times\vec{n}$. This correction to the magnetic field results in a net field $\vec{B}_\Sigma = \vec{B}_R + \vec{B}_\bot$ which is different to $\vec{B}_R$. Considering the adiabatic limit, where $|\vec{B}_R| \gg |\vec{B}_\bot |$, we recovered exactly the results of Sinova \emph{et al.} describing the SHE in the Rashba system. The field $\vec{B}_\bot$ was shown to be described quantum mechanically by a gauge field in time space. In the Luttinger system, the adiabatic assumption results in a non-trivial momentum space curvature which enters the equation of motion as the spin-dependent anomalous velocity component \eqref{luttanom.eq}.

Thus, at first glance it appears that the two effects are rather independent phenomena. However, an interesting duality exists between the two effects. In the former Luttinger case, a \emph{spin-dependent anomalous velocity} pushes opposite spin species to opposite lateral sides of the sample. The magnetic field responsible for this effect is the Berry curvature defined by the $\vec{k}$-space gauge field. On the other hand, in the Rashba system, a \emph{momentum-dependent magnetic field} polarizes electrons along opposite directions out-of-the-plane depending on their transverse propagation direction. The magnetic field responsible for this effect is defined by the $t$-space gauge field. Given this duality, it would be tempting to ask whether there is any underlying relation between the two pictures.

We proceed to consolidate the link between the two effects by investigating the connection between the anomalous velocity due to the Berry curvature, and the presence of the $\vec{B}_\bot$ term. In this endeavour, we employ the reciprocal space analogue of the analysis by Aharanov and Stern\cite{stern} of the origin of the Berry's curvature in real space. We consider again the general spin-orbit Hamiltonian in Eq.\ \eqref{sochamil.eq}. The velocity along the $i$-th coordinate is given by Hamilton's equation:
\begin{eqnarray}
v_i &=& \frac{1}{i \hbar}\left[r_i,\mathcal{H}\right], \nonumber\\
&=& \frac{p_i}{m}-\gamma \frac{\partial \vec{B}(\vec{k})}{\partial p_i}\cdot\vec{\sigma}.
\label{veli.eq}
\end{eqnarray}
When the magnetic field is time-dependent, the spins see an additional magnetic field $\vec{B}_\bot$. Assuming that spins align to $\vec{B}_\Sigma = \vec{B}+\vec{B}_\bot$, the spin vector, to first order, is given by $\vec{\sigma}=\vec{B}_\Sigma / |\vec{B}_\Sigma|$. Writing the spin-orbit field $\vec{B} = |\vec{B}|\vec{n}$, the partial derivative in Eq.\ \eqref{veli.eq} can be expanded into its magnitude and directional parts as $(\partial |\vec{B}|/\partial p_i)\vec{n} + |\vec{B}|\partial \vec{n}/\partial p_i$. Taking the adiabatic limit $|\vec{B}_\Sigma|/|\vec{B}|\rightarrow 1$, the second term in the velocity expression becomes
\begin{eqnarray}
v_i &=&  -\gamma\frac{\partial |\vec{B}|}{\partial p_i}-\frac{\hbar}{2}\left(\dot{\vec{n}}\times\vec{n}\right)\cdot\frac{\partial \vec{n}}{\partial p_i}.
\end{eqnarray}
Writing $\dot{\vec{n}}=\dot{k_j} \partial \vec{n}/\partial k_j$, where the summation over $j$ is implicit, and rearranging the terms we then get
\begin{equation}
v_i =  -\gamma\frac{\partial{|\vec{B}|}}{{\partial p_i}}-\frac{\dot{k_j}}{2}\left( \frac{\partial \vec{n}}{\partial k_i}\times \frac{\partial \vec{n}}{\partial k_j} \right)\cdot\vec{n}.
\label{anom-vel.eq}
\end{equation}
The first term in the above equation represents a velocity term that is due to the inhomogeneity of the spin-orbit field $\vec{B}$ in momentum space, i.e.\ it is the reciprocal space analogue of the Stern-Gerlach force. Remarkably, the second term in Eq.\ \eqref{anom-vel.eq} is the anomalous velocity of electrons due to Berry's curvature in $\vec{k}$-space. This becomes clearer when written in terms of the magnetic field vector $\vec{B}=|\vec{B}|\vec{n}$, 
\begin{eqnarray}
v_i^{\text{anom.}}&=&-\dot{k_j} \frac{\vec{B}}{2|\vec{B}|^3}\cdot\left( \frac{\partial \vec{B}}{\partial k_i}\times \frac{\partial \vec{B}}{\partial k_j} \right),\nonumber\\
&=& - \epsilon_{ijk}\dot{k_j} \Omega_k (\vec{k}).
\end{eqnarray}
We find that this is exactly the anomalous velocity component in Eq.\ \eqref{eom.eq}, which arises from the Berry's curvature in $\vec{k}$-space. By letting the spin vector to be aligned anti-parallel to the net magnetic field, $\vec{\sigma}=-\vec{B}_\Sigma / |\vec{B}_\Sigma|$ in Eq.\ \eqref{veli.eq}, the correct spin-dependence of the anomalous velocity can be obtained.
Thus, we have shown that the anomalous velocity due to the Berry curvature actually arises because of the $\vec{B}_\bot$ magnetic field component, which in turn is related to the time component of the gauge field. This gauge field component therefore plays an equally important role in the SHE in the Luttinger system, as it does in the Rashba one, and acts as the unifying bridge between the two effects.
\section{Summary}
The primary motivation for this paper is to establish the link between the two intrinsic SHEs reported in Refs.\ [\onlinecite{murakami,sinova}], which has not been clarified hitherto. We first considered the intrinsic SHE in the Luttinger system, which is driven by the spin-dependent anomalous velocity due to the non-trivial curvature of momentum space. However, this theoretical picture is not applicable in the planar Rashba system. Instead, the SHE in the Rashba system was shown to arise from spins acquiring a component (in the adiabatic sense) along an additional effective magnetic field $\vec{B}_\bot$, arising from the time-dependence of the system. This field component was shown to be described by a gauge field in time space. Finally, we showed that in the adiabatic limit, $\vec{B}_\bot$ is also the origin of the anomalous velocity due to the momentum space Berry curvature. Thus, we conclude that the intrinsic SHEs in the two systems are simply different manifestations of $\vec{B}_\bot$, and that this term provides a unifying link between the two effects.
\appendix*
\section{Classical derivation of effective magnetic field component, $\dot{\vec{n}}\times\vec{n}$}
\begin{figure}[!ht]
\centering
\resizebox{0.75\columnwidth}{!}{
\includegraphics{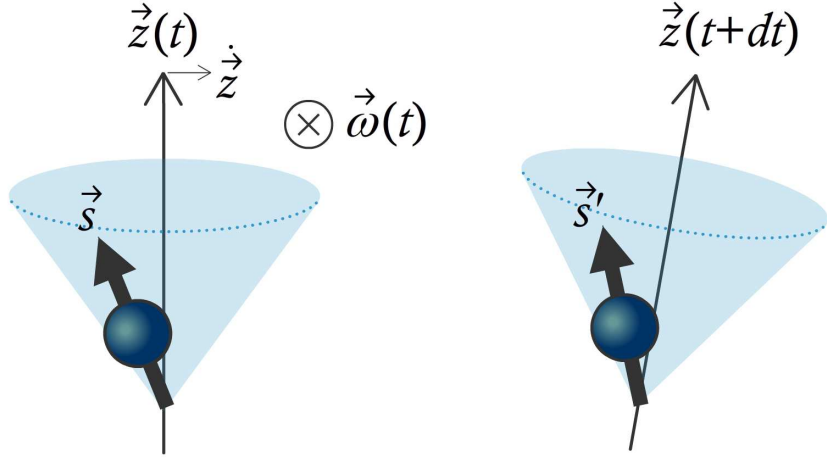}}
\caption{(Color online) (left) The classical spin vector $\vec{s}(t)$ precesses about the magnetic field which is along the $\vec{z}$ direction at some instant $t$. Because of the time-dependence of the magnetic field, the spin is also subject to a rotation about $\vec{\omega}(t)=\dot{\vec{z}}\times\vec{z}$ (see text) which transforms it from the frame at time $t$ (left) to the frame at time $t+dt$ (right). The $\vec{\omega}(t)$ acts as an additional magnetic field which governs the overall spin dynamics.}
\label{spin_prec.fig}
\end{figure}

Consider the dynamics of the spin vector $\vec{s}(t)$ in a time-dependent magnetic field $\vec{B}(t)$, 
\begin{equation}
\dot{\vec{s}}(t) = g\left(\vec{s}(t)\times\vec{B}(t)\right),
\end{equation}
where $g$ is the coupling factor. To solve the above equation, we freeze the time-dependence by transforming to a rotated coordinate frame at each point in time, such that the $\tilde{z}$-axis is aligned with the magnetic field. 
A spin vector $\vec{s}$ defined relative to the coordinate frame at time $t$, is expressed as the vector $\vec{s}^{\ '} = \vec{s} + \vec{s}\times\vec{\omega}(t)dt$ in the coordinate frame at time $t+dt$, where $\vec\omega (t)$ is the generator of infinitesimal rotations (see Fig.\ \ref{spin_prec.fig}).
The choice of $\vec\omega(t)$ is not unique; however, specifically choosing $\vec\omega(t)=\dot{\vec{z}}\times\vec{z}$ where $\vec{z}$ is the unit vector $\vec{n}=\vec{B}/|\vec{B}|$ as seen in the rotated frame, coincides with the parallel transport of the coordinate frames.\cite{stern,anandan} Suppose we have a vector representing the spin, $\vec{s}(t)$, in the rotated frame at time $t$. At time $t+dt$, this vector becomes [relative to frame $t+dt$] $\vec{s}(t+dt)+\vec{s}(t+dt)\times\vec\omega(t)dt$ where $\vec{s}(t+dt)\approx\vec{s}(t)+g (\vec{s}(t)\times|\vec{B}|\vec{z}) dt$. For infinitesimally small $dt$, we may write $\vec{s}(t+dt)$ [in frame $t$] $\approx \vec{s}(t+dt)+\vec{s}(t+dt)\times\vec\omega(t)dt$ [in frame $t+dt$]. The right-hand-side of the resulting equation becomes $\vec{s}(t)+g(\vec{s}\times|\vec{B}|\vec{z})dt + \vec{s}\times\vec\omega(t)dt + \mathcal{O}(dt^2)$. Rearranging, and taking the limit $dt\rightarrow 0$, we have
\begin{eqnarray}
\lim_{dt\rightarrow 0} \frac{\vec{s}(t+dt)-\vec{s}(t)}{dt} \equiv \dot{\vec{s}} &=& g\left(\vec{s}\times|\vec{B}(t)|\vec{z}\right) + \vec{s}\times\vec\omega(t),\nonumber\\
&=&\vec{s}\times\left(g|\vec{B}(t)|\vec{z}+\dot{\vec{z}}\times\vec{z}\right).
\end{eqnarray}
Therefore, as seen in the laboratory frame, there is an additional, effective magnetic field $\vec{B}_{\text{eff.}}=\vec{B}(t)+g^{-1}\dot{\vec{n}}\times\vec{n}$.

\end{document}